\documentclass[preprint2]{aastex}  % 2-column preprint
\usepackage{natbib}\usepackage{txfonts}\usepackage{balance}
\usepackage{graphicx}

\shorttitle{Solar mass loss, accretion, and overshoot}
\shortauthors{}

\begin{document}

\title{Exploring mass loss, low-Z accretion, and convective overshoot in solar models to mitigate the solar abundance problem}

\author{Joyce Ann Guzik and Katie Mussack}
\affil{
XTD-2,
MS T-086, Los Alamos National Laboratory,
Los Alamos, New Mexico, 87545
}
\email{joy@lanl.gov}

%\authorrunning{Guzik and Mussack}
%\titlerunning{Mass loss, accretion, and convective overshoot}

\begin{abstract}

Solar models using the new lower abundances of Asplund et al. (2005, 2009) or Caffau et al. (2008, 2009) do not agree as well with helioseismic inferences as models that use the 
higher Grevesse \& Noels (1993) or Grevesse \& Sauval (1998) abundances. Adopting the new abundances leads to models with sound speed discrepancies of up to 1.4$\%$ below the base of the convection zone (compared to discrepancies of less than 0.4$\%$ with the old abundances), a convection zone that is too shallow, and a convection zone helium abundance that is too low. Here we review briefly recent attempts to restore agreement, and we evaluate three  changes to the models: early mass loss, accretion of low-Z material, and convective overshoot. One goal of these attempts is to explore models that could preserve the structure in the interior obtained with the old abundances while accommodating the new abundances at the surface. Although the mass-losing and accretion models show some improvement in agreement with seismic constraints, a satisfactory resolution to the solar abundance problem remains to be found. In addition, we perform a preliminary analysis of models with the Caffau et al. (2008, 2009) abundances that shows that the sound speed discrepancy is reduced to only about 0.6$\%$ at the convection zone base, compared to 1.4$\%$ for the Asplund et al. (2005) abundances and 0.4$\%$ for the Grevesse \& Noels (1993) abundances. Furthermore, including mass loss in models with the Caffau et al. (2008, 2009) abundances may improve sound speed agreement and help resolve the solar lithium problem.

\end{abstract}


\section{Introduction}
\label{sect:intro}

\subsection{The Solar Abundance Problem}
\label{sect:abundprob}

Before 2005, we believed we knew very well how to model the evolution and interior structure of the sun.  Evolved solar models with the latest input physics (including diffusive helium and element settling and without tachocline mixing) reproduced the sound speed profile determined from seismic inversions to within 0.4$\%$, as well as the seismically-inferred convection zone depth and convection zone helium abundance. Solar interior modelers had little impetus to progress beyond one-dimensional spherical models of the sun, with perhaps the exception of introducing some additional mixing below the convection zone (hereafter CZ) to deplete surface lithium and reduce the small remaining sound-speed discrepancy at the CZ base.  For the \citet[][hereafter GN93]{GN93} or \citet[][hereafter GS98]{GS98} abundances, the simplest `spherical sun' assumptions appeared nearly adequate for solar modeling. These include one-dimensional zoning (concentric shells in hydrostatic equilibrium), initial homogeneous composition, negligible mass loss or accretion, neglecting rotation and magnetic fields, simple surface boundary conditions, mixing-length theory of convection (e.g., \citet{Bohm_1958}), and no additional mixing or structural changes from convective overshoot, shear from differential rotation, meridional circulation, waves, or oscillations.

However, new analyses of solar spectral lines revise downward the abundances of elements heavier than hydrogen and helium, particularly the abundances of carbon, nitrogen, and oxygen that contribute to the opacity just below the CZ. See Table \ref{table:abund} for a summary of some of the major abundance revisions over the last twenty years.  

\begin{table*}
\caption{Mass fractions of metals in the present-day photosphere, Z, and ratio of metals to hydrogen mass fraction, Z/X, evaluated over the last twenty years.}
\label{table:abund}
\begin{center}
\begin{tabular}{llll}
\hline
Year & Source & Z & Z/X  \\
\hline
1989 & \citet{AG89}   &$ 0.0201 $ & $0.0274 $ \\
1993 & \citet{GN93}   &$ 0.0179 $ & $0.0244 $ \\
1998 & \citet{GS98}   &$ 0.0170 $ & $0.0231 $ \\
2005 & \citet{AGS05}  &$ 0.0122 $ & $0.0165 $\\
2009 & \citet{AGSS09} &$ 0.0134 $ & $0.0181 $\\
2009 & \citet{Ludwig_2009} &$0.0154 $ & $0.0209 $\\
\hline
\end{tabular}
\tablenotetext{a}{Values for Z are inferred from Z/X assuming Y$=$0.248.}
\tablenotetext{b}{Uncertainties are $<$10$\%$.}
\end{center}
\end{table*}

Compared to the older GS98 and GN93 abundances, the \citet[][hereafter AGS05]{AGS05} abundance of C is lower by 35$\%$, N by 27.5$\%$, O by 48$\%$, and Ne by 74$\%$.  The abundances of elements from Na to Ca are lower by 12 to 25$\%$, and Fe is decreased by 12$\%$.  For the GS98 abundances, the ratio of the element to hydrogen mass fraction Z/X = 0.023, and the heavy element abundance Z $\sim$ 0.018, while, for the new abundances, Z/X = 0.0165, and Z$\sim$0.0122.  Models evolved with the AGS05 abundance mixture give worse agreement with helioseismic constraints; the sound-speed discrepancy is 1.4$\%$ below the CZ base, and the CZ depth is shallow and CZ helium abundance is low compared to those derived from seismic inversions.

Recently, \citet[][hereafter AGSS09]{AGSS09} re-evaluated the spectroscopic abundances, carefully considering the atomic input data and selection of spectral lines and using improved radiative transfer and opacities. They revised the heavy element abundance to Z/X = 0.0181 and Z = 0.0134. This slight increase over the AGS05 values yields some improvement in agreement with seismic constraints \citep[see][]{Serenelli_2009}, although the higher abundances of GN93 or GS98 still provide the best agreement. In addition, the Cosmological Impact of the FIrst STars Team and its collaborators used the 3D model atmosphere code CO$^5$BOLD to perform an independent investigation of the solar abundances  \citep[][hereafter CO$^5$BOLD]{Caffau_2008a,Caffau_2008b,Caffau_2009,Ludwig_2009}. They derived heavy element abundances of Z/X = 0.0209 and Z = 0.0154, in between the AGSS09 and GN93 values.

Spectroscopic determinations measure the photospheric abundances. The continuous convective overshoot into the photosphere should leave the photosphere with the same abundances as the convection zone. In addition, convective timescales are much shorter than element diffusion and evolution timescales, making the convection zone well mixed and homogeneous. Therefore the spectroscopic abundances should be indicative of the abundances throughout the convection zone. However, we are reluctant to dismiss the recent abundance re-analyses because of the many improvements in the physics and models included, namely 3D dynamical atmosphere models, non-local thermodynamic equilibrium corrections for important elements, and updated atomic and molecular data.  Line profile shapes now agree nearly perfectly with observations. Also, it is impressive that abundances derived from several different atomic and molecular lines for the same element now are consistent. 

For this paper, we will focus on the AGS05 abundances as the most extreme example of the lower abundance determinations. However, we will devote section \ref{sect:COBOLD} to a preliminary exploration of the CO$^5$BOLD abundances.
We first review the results of helioseismic tests using the old and new abundances and ongoing attempts to resolve these discrepancies (Section \ref{sect:intro}). We discuss in detail the following three mitigation attempts: an early mass-loss phase in solar evolution (Section \ref{sect:massloss}), accretion of low-Z material early in the sun's lifetime (Section \ref{sect:accret}), and extending the CZ below the depth inferred by helioseismology (Section \ref{sect:overshoot}). We then examine including some of these adjustments in models with the higher metallicity of CO$^5$BOLD (Section \ref{sect:COBOLD}). The models used to explore all of these changes are described in Section \ref{sect:models}.

\subsection{Helioseismology and solar models}
\label{sect:mod_helio}

Helio- and asteroseismology have turned out to be excellent tools to test the physics of stellar models. \citet{BA04a, BP04, Turck_2004} authored some of the first papers to examine the effects of the new lower abundances on solar models, demonstrating that the new abundances lead to greater discrepancy with seismic inferences.  This is demonstrated in Table \ref{table:ZYR} which compares our calibrated evolution models using the GN93 and AGS05 mixtures (details of our models can be found in Section \ref{sect:models}).  For the AGS05 model, the CZ helium abundance Y is low (0.22730) and the CZ base  (R$_{\rm{CZB}}$ = 0.72944 R$_{\odot}$) is shallow compared to the seismically inferred CZ Y abundance of 0.248$\pm$0.003 and CZ base radius of 0.713$\pm$0.001 R$_{\odot}$ from \citet{BA04a}. 

\begin{comment}
\begin{table}
\caption{Properties of our calibrated standard solar models \citep{GWC04,GWC05}.}
\label{table:standardmodels}
\begin{center}
\begin{tabular}{lll}
\hline
Model Property & GN93 Mixture & AGS05 Mixture  \\
\hline
Y$_{\rm{o}}$ &$ 0.26930 $ & $0.25700 $ \\
Z$_{\rm{o}}$   &$ 0.0197 $ & $0.01350 $ \\
$\alpha$  &$ 2.0379 $ & $2.0004 $ \\
Z$_{\rm{CZ}}$  &$  $ & $ $\\
Y$_{\rm{CZ}}$   &$ 0.24124 $ & $0.22730 $\\
R$_{\rm{CZB}}$ (R$_{\odot}$ ) &$ 0.71254 $ & $0.72172 $
\\
\hline
\end{tabular}
\end{center}
\end{table}
\end{comment}

\begin{figure}[t!]
%\resizebox{\hsize}{!}{\includegraphics{sound.eps}}
\resizebox{\hsize}{!}{\includegraphics{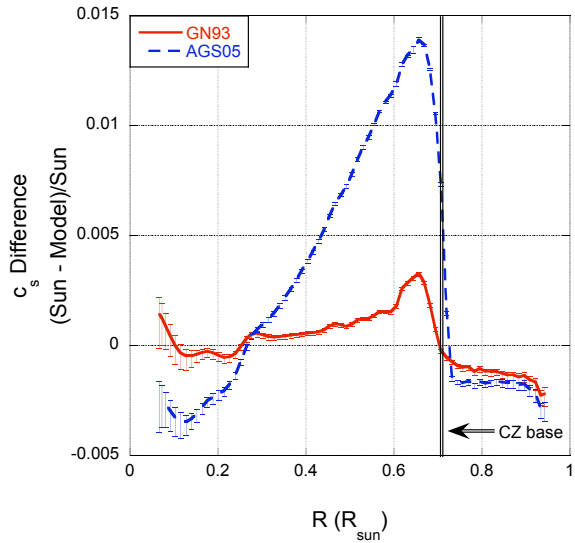}}
\caption{\footnotesize Difference between inferred and calculated sound speeds with error bars for models with the GN93 and AGS05 abundances. The sound speed inversion is from \citet{BPB00}. The seismically inferred convection zone base at R = 0.713 R$_{\odot}$ \citep{BA04a} is shown with the vertical line.}
\label{fig:c_standard}
\end{figure}

\begin{figure}[t!]
%\resizebox{\hsize}{!}{\includegraphics{omincScott.eps}}
\resizebox{\hsize}{!}{\includegraphics{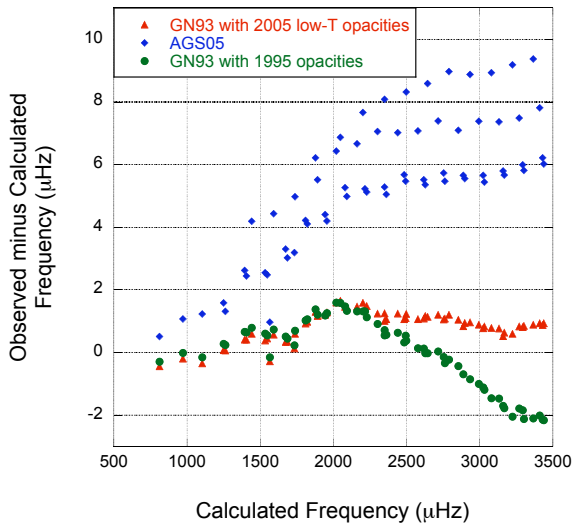}}
\caption{\footnotesize Observed minus calculated versus calculated frequency for our models with the GN93 mixture (triangles or circles), and the AGS05 mixture (diamonds) for modes of degree $\ell$ = 0, 2, 10, and 20. The models producing the triangle and diamond points use the newer \citet{Fer05} low-temperature opacities, while the model with the circle points uses 1995 Alexander \& Ferguson opacities. The calculated frequencies were computed using the \citet{Pesnell_1990} non-adiabatic stellar pulsation code, and the data are from \citet{Chaplin_2007,ST96,Garcia_2001}.}
\label{fig:O-C_standard}
\end{figure}

Figure \ref{fig:c_standard} shows the differences between inferred and calculated sound speed for calibrated evolved models using the old and new abundances. The inferred sound speed is from \citet{BPB00}. The uncertainties in sound speed inversions are much smaller than the differences between these curves, at most a few widths of the plotting line.  Figure \ref{fig:O-C_standard} shows the observed minus calculated frequency differences for modes of angular degree $\ell$ = 0, 2, 10, and 20 that propagate into the solar interior below the convection zone. The calculated frequencies were computed using the \citet{Pesnell_1990} non-adiabatic stellar pulsation code. The observational data are from BiSON \citep{Chaplin_2007}, LowL \citep{ST96}, or GOLF \citep{Garcia_2001}. The observational uncertainties for the modes are less than 0.1 $\mu$Hz, much smaller than the model discrepancies. The O-C trends for the model with the \citet{Fer05} low-temperature opacities are flatter for higher frequencies that are more sensitive to the solar surface. These newer opacities include an improved treatment of grains, finer wavelength spacing, and additional molecular lines, and are higher by 12$\%$ up to a factor of three. As illustrated by both of these plots, the discrepancy with the new abundances is much larger than with the old abundances. 

Helioseismology has also been used to investigate heavy element abundances. \citet{Lin_2005} find that a reduction in carbon abundance, in the direction of the AGS05 abundances, can improve sound speed inference. However, \citet{Lin_2007} show that lower Z increases the discrepancy with adiabatic index inversions; \citet{AB06} use the ionization signature in the sound speed derivative to infer Z$_{\rm{CZ}}$ = 0.0172$\pm$0.002. These results favor the old, higher abundances. In addition, \citet{Chaplin_2007} use small frequency separations between low-degree modes that are sensitive to the core structure to constrain the core Z abundance. They find Z$_{\rm{core}}$ = 0.0187-0.0239. \citet{Zaatri_2007} find that the mean low-degree frequency spacings of a model using AGS05 abundances are incompatible with those determined from the GOLF measurements of \citet{Lazrek_2007} and \citet{Gelly_2002}. Similarly, \citet{Basu_2007} find that models constructed with low metallicity are incompatible with the small frequency spacing and frequency separation ratios calculated from BiSON data \citep{Chaplin_1996}. These results do not rule out accretion, enhanced diffusion, or other options that can retain high core Z. However, they do disfavor the prospect that the new, lower abundances were present initially throughout the sun.

\subsection{Attempts to restore agreement through solar model modifications}
\label{sect:attempts}

Here we briefly review recent attempts to adjust solar models in order to mitigate the discrepancy with seismic constraints for the new abundances. For a more detailed review of previous attempts, see, for example, \citet{BA08}.  In the following sections, we discuss increased opacities below the CZ ($11-30\%$); increased neon abundance ($\times$$\sim$4); increased abundances (within uncertainty limits, or using alternative determinations); enhanced diffusive settling rates ($\times$1.5 or more); accretion of lower-Z material early in the sun's lifetime; structure modification below the CZ base due to radiative damping of gravity waves; tachocline mixing (also used with old abundances); convective overshoot; and combinations of the above. The conclusion has been that it is difficult to match simultaneously the new Z/X and helioseismic constraints for CZ depth, sound speed and density profiles, and CZ helium abundance by applying these changes.

\subsection{Opacity increases}
\label{sect:op_inc}

Heavy-element abundances primarily affect solar structure through their effect on opacity, which affects the structure of the radiative zone and the location of the CZ base. The structure of most of the convection zone is essentially independent of the opacity. \citet{JCD09} determined the change in opacity required to restore the sound-speed agreement of a solar model using the AGS05 abundances to the level of success originally attained with the GN93 abundances. They find that opacity would need to be increased ranging from about 30$\%$ below the CZ to a few percent in the core.  

Although improvements to the solar model can be made by {\it ad~hoc} opacity increases, there is little justification for such large enhancements. The presently available opacity tables from three separate projects (OPAL, OP, and LANL T-4) for conditions below the CZ differ by only a few percent \citep{Neuforge_2001,Badnell_2005}, making it difficult to justify such large opacity enhancements. Using the Los Alamos National Laboratory T-4 opacity library data \citep{Magee_1995, Hakel_2006}, \citet{GKK09} find that, to obtain a 30$\%$ opacity increase with the new abundances, the contribution from oxygen alone would need to increase by a factor of two to three.  Alternatively, the iron absorption contribution would need to increase by a factor of three.

Including additional elements has a negligible effect on Rosseland mean opacities for solar interior conditions.  The Lawrence Livermore OPAL opacities for the AGS05 mixture included 17 of the most abundant elements. With the LANL T-4 opacity library, \citet{GKK09} find that including in the mixture all of the elements up to atomic number Z=30 increases the mixture opacity by only 0.2$\%$ for solar interior conditions.  Including additional elements in the mixture from 30 $<$ Z $<$ 93, an 83-element mixture, further increases the mixture opacity by less than 0.1$\%$. An as-yet unidentified error in calculations of the line wings for K-shell transitions in O, C, N, and Ne at energies around 800 eV, the peak of the weighting of the Rosseland mean opacity for the temperature conditions at the CZ base, could provide some increase in the calculated Rosseland mean opacities \citep{GKK09}.

In order to shed light on the issue, \citet{Turck_2009} suggest experimental investigations of opacity coefficients for the radiative zones of solar-like stars. The coming large laser facilities (LIL +PETAL, OMEGA EP, FIREX II, LMJ, NIF) have the potential to attain sufficiently high temperatures and high densities at LTE along with the precise diagnostics required for stellar opacity measurements. Experiments to investigate the opacities relevant for stellar interiors are being conducted at Sandia's Z facility by \citet{Bailey_2009}.

\subsection{Neon and other element abundance increases}
\label{sect:abund_inc}

For a while, it was thought that an increase in the solar neon abundance provided the most plausible resolution to this problem.  Neon is not measured in the photosphere due to a lack of suitable spectral lines. Instead, its abundance is determined relative to oxygen using lines formed in the solar corona, XUV and gamma ray spectroscopy of quiet and active regions, and solar wind particle collections. AGS05  adopt a Ne/O abundance ratio of 0.15, and apply this ratio to the photospheric oxygen abundance to derive the neon abundance. The neon abundance has been revised downward by $74\%$ from the GS98 value, for the most part due to the oxygen abundance reduction.

Several groups explored solar models with enhanced neon. Some improvement in agreement was shown for Ne increases ranging from 0.5-0.67 dex \citep{AB05, BBS05, TCP05, Zaatri_2007, DP06}. However, \citet{Lin_2007} find that increasing Ne alone actually increases the discrepancy in the adiabatic exponent in the region 0.75-0.9 R$_{\odot}$. They find that the discrepancy is reduced if only C, N, and O abundances are increased.

More modest Ne enhancements, combined with increases in the other element abundances of $\sim$0.05 dex, at the limit of the AGS05 uncertainties, have also been considered.  The best model of \citet{BBS05} has Ne enhanced by 0.45 dex (2.8$\times$), Ar by 0.4 dex, and C, N, and O by 0.05 dex.  This model produces reasonably good agreement with the inferred sound speed and density profile, and has CZ base radius 0.715 R$_{\odot}$ and acceptable CZ Y= 0.2439. Of course any increase in abundances from the AGS05 value will mitigate the problem by increasing opacities below the CZ. For examples, see \citet{Zaatri_2007} or \citet{Turck_2004}.

\subsection{Enhanced diffusion}
\label{sect:diff}

Several groups, e.g., \citet{BA04a}, \citet{Montalban_2004}, \citet[][hereafter GWC05]{GWC05}, and \citet{YB07} considered the effects of enhanced diffusion.  At first this idea might seem promising, because the solar interior could have higher abundances that give good sound speed agreement, while the CZ elements could be depleted to the AGS05 photospheric values.  In practice, the required diffusion increases are quite large (factors of 1.5 to 2 on absolute rates), and enhanced diffusion also depletes the CZ Y abundance to well below the seismic determination and leaves the CZ too shallow.

GWC05 investigate the enhancement of thermal diffusion for elements and He by different amounts. A model with resistance coefficients $\times$1/4 for C, N, O, Ne, and Mg and $\times$2/3 for He shows some improvement in the sound-speed discrepancy, but the CZ depth is still a little shallow (0.718 R$_\odot$), and the CZ Y is still a little low (0.227). Moreover, there is no justification for these {\it ad~hoc} changes in thermal diffusion coefficients.The diffusion coefficients themselves should not be in error by such a large factor.

\subsection{Gravity waves and dynamical effects}
\label{sect:g_waves}

Arnett, Meakin, \& Young (2006, private communication) have been investigating, following \citet{Press_1981} and \citet{PR81}, the effects of gravity waves excited and launched inward at the CZ base. The radiative damping of these waves as they travel inward deposits energy and changes the solar structure in the same way as would an opacity enhancement. The amount of damping and distance that the waves propagate depend on the initial amplitudes and the degree of the mode, with low-frequency, high-degree waves damped more heavily, after traveling a shorter distance. The expected wave spectrum and amplitudes still need to be worked out, but could remove as much as one third of the sound-speed discrepancy. More recently, \citet{Arnett_2009} have re-examined convection at the surface and sub-surface layers of the sun, proposing a way to eliminate astronomical calibration from stellar convection theory. By choosing characteristic lengths that are determined by the flow, they eliminate the need for the free parameters traditionally used in mixing length theory. They show that some of the discrepancy between the new abundances and helioseismic inferences may result from the neglect of hydrodynamic processes in the standard solar model.

\subsection{Combinations of effects}
\label{sect:combos}

In addition to single changes to solar models, several groups considered combinations of changes, such as diffusion, opacity, and abundance enhancements. See, for example, \citet{BA04a}, \citet{Montalban_2004}, and \citet{BSB06}. 

Although the above modifications to the input physics have achieved some success in restoring agreement between seismic constraints and models that use the new abundances, agreement is not fully restored and there is little physical justification for the proposed changes. In this paper, we discuss the motivation for and the results of three additional attempts to restore agreement: an early mass loss phase, accretion of low-Z material, and convective overshoot.

\section{Solar evolution models}
\label{sect:models}

Solar models require input data for opacities such as OPAL \citep{IR96} or OP \citep{SB04} supplemented by low-temperature opacities \citep[e.g.,][]{Fer05}; equation of state such as OPAL \citep{RSI96}, MHD \citep{MHD}, or CEFF \citep{CEFF}; nuclear reaction rates \citep[e.g.,][]{Angulo_1999}; and diffusive element settling \citep[e.g.,][]{Burgers_1969, CGK89, TBL94}.  

The solar models produced by the Los Alamos group shown hereare evolved from the pre-main sequence using an updated version of the one-dimensional evolution codes described in \citet{Iben_1963, Iben_1965a, Iben_1965b}. The evolution code uses the SIREFF EOS  \citep[see][]{GS97}, Burgers' diffusion treatment as implemented by \citet{CGK89}, the nuclear reaction rates from \citet{Angulo_1999} with a correction to the $^{14}$N rate from \citet{Formicola_2004}, and the OPAL opacities \citep{IR96} supplemented by the \citet{Fer05} or Alexander \& Ferguson (private communication, 1995) low-temperature opacities. The \citet{Fer05} low-temperature opacities include an improved treatment of grains, finer wavelength spacing, and additional molecular lines and are higher than the Alexander \& Ferguson (1995) low-temperature opacities by 12$\%$ up to a factor of three. As discussed in Section \ref{sect:mod_helio}, the observed minus calculated frequency trends for a model with the newer low-T opacities are flatter for higher frequencies that are sensitive to the solar surface, as seen in Figure \ref{fig:O-C_standard}. 

The models are calibrated to the present solar radius 6.9599$\times10^{10}$ cm \citep{Allen_1973}, luminosity 3.846$\times 10^{33}$ erg/g \citep{Willson_1986}, mass 1.989$\times 10^{33}$  g \citep{CT86}, age 4.54$\pm$0.04 Gyr \citep{Guenther_1992}, and adopted photospheric Z/X ratio.  For evolution models, the initial helium abundance Y, initial element mass fraction Z, and mixing length to pressure-scale-height ratio $\alpha$ are adjusted so that the final luminosity, radius, and surface Z/X match the observational constraints to within uncertainties. See \citet{GWC05} for references and a description of the physics used in the evolution and pulsation codes and models. 

\begin{comment}
 Note for future use:
  luminosity 3.8418$\times 10^{33}$ erg/g \citep{Frohlich_1998, Bahcall_1995}
  age 4.57 Gyr \citep{Bahcall_1995}
%\bibitem[{Bahcall et al.(1995)}]{Bahcall_1995}
%Bahcall, J.N., Pinsonneault, M.H., \& Wasserburg, G.J.  1995, Rev. Modern Phys., 67, 781
%\bibitem[Fr\"{o}hlich \& Lean (1998)]{Frohlich_1998}
%Fr\"{o}hlich, C., \& Lean, J. 1998, Geophys.Res. Lett., 25, 4377
\end{comment}

\section{Mass loss}
\label{sect:massloss}

\subsection{Motivation and method}
\label{sect:massloss_meth}

\citet{Willson_1987} explored the possibility that significant mass loss could occur during the early part of the main-sequence for $\sim$1-2.5 M$_{\odot}$ stars. They considered mass-loss rates ranging from $10^{-9}$ to $10^{-8}$ M$_{\odot}$/yr that would remove a substantial fraction of mass from a star before it evolves off the main sequence. Mass loss in these stars, possibly including the early sun, would be driven by pulsation, which provides the necessary mechanical energy flux, and facilitated by rapid rotation. The mass-loss rate would diminish upon the development of a surface convection zone, which channels mechanical energy away from pulsation, and of magnetic fields which provide angular momentum transfer and rotational braking. \citet{GWB87} showed that mass-losing solar models have steeper molecular weight gradients, shorter main-sequence lifetimes, higher $^8$B neutrino fluxes, deeper surface convection, higher surface $^3$He abundances, and earlier, more pronounced dredge-up of CN cycle processed material in the post-main-sequence phase compared to the standard model. In addition, mass-losing models predict complete destruction of protosolar Li and Be, requiring a mechanism, such as production in spallation reactions or flares, for partial replenishment to the observed surface abundances.

Such mass loss in other stars could potentially explain blue stragglers and the earlier-than-predicted dredge-up of carbon and nitrogen in solar-mass stars ascending the first red giant branch \citep[see][]{Guzik_1988}. However, there are also drawbacks. If the sun remains at too high a mass for too long, all surface Li in the subsequent solar model at present age is destroyed, too much surface ${}^3\rm{He}$ is produced, and discrepancies with the inferred sound speed arise \citep{Guzik_1988}. In addition, \citet{GC95} compare observed and calculated $p$-mode oscillation frequencies to test the structure of solar models including early main-sequence mass loss. They show that extreme solar mass loss has a significant effect on solar structure and can be ruled out by the $p$-mode oscillation frequencies.

While more extreme early main-sequence mass loss has not been 
observed, smaller mass loss in the sun of about 0.1 M$_\odot$ looks promising to solve a number of problems. The advantages of an early mass loss phase in solar evolution include: solving the faint early sun problem, explaining early liquid water on Mars, early inner solar system bombardment, and solar lithium destruction. 

Models with early mass loss using the older, higher element abundances were explored previously by \citet{GWB87}, \citet{SF92}, \citet{GC95}, and \citet{SB03}. In addition, \citet{MM07} recently assessed consequences for Earth climate and solar system formation. Here we re-investigate solar mass loss in light of the new abundances. In our models, we use the mass-loss treatment implemented by \citet{Brunish_1981} in the \citet{Iben_1963, Iben_1965a, Iben_1965b} code.
We evolve two models with initial masses 1.3 and 1.15 $\rm{M}_{\odot}$, having exponentially-decaying mass-loss rates with an e-folding time of 0.45 Gyr. Following \citet{GWB87}, we adopt this exponential mass-loss prescription  because it is simple and decreases smoothly with time. This is a physically plausible description, as the mass-loss rate should be highest when a rotating star arrives on the main sequence within the instability strip, where pulsation and rotation can facilitate mass loss. The mass loss should then decrease as the star moves out of the instability strip, ceases to pulsate, spins down, and develops a surface convection zone. The present solar mass-loss rate is 2 $\times10^{-14}$ $\rm{M_{\odot}/yr}$ \citep[e.g.,][]{Feldman_1977}, too small to affect the sun's evolution.  The initial mass-loss rates for the two models are 6.55 and 3.38 $\times10^{-10}$ M$_{\odot}/$yr, respectively.

\subsection{Results}
\label{sect:massloss_results}
 
\begin{figure}[t!]
% \resizebox{\hsize}{!}{\includegraphics{LuminosityvstimeMLmodels.eps}}
 \resizebox{\hsize}{!}{\includegraphics{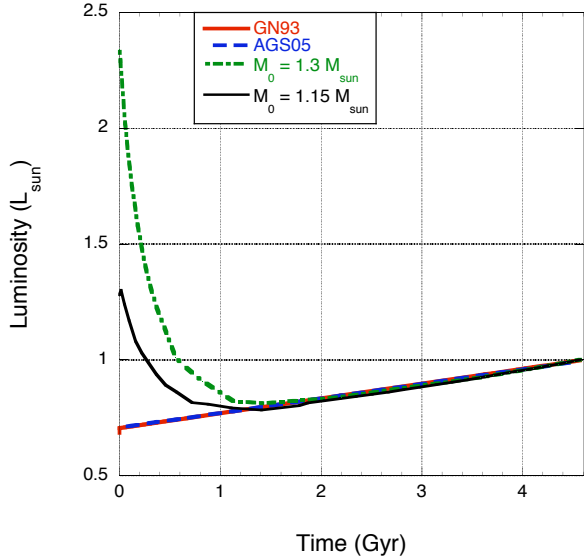}}
 \caption{\footnotesize Luminosity versus time for standard one solar-mass models using the GN93 and AGS05 abundances and for two mass-losing models using the AGS05 abundances with initial mass 1.3 and 1.15 M$_{\odot}$.  Mass-loss rates are exponentially decaying with e-folding time 0.45 Gyr.}
 \label{fig:lum_massloss}
\end{figure}
 
\begin{figure}[t!]
% \resizebox{\hsize}{!}{\includegraphics{SoundMassLoss.eps}}
 \resizebox{\hsize}{!}{\includegraphics{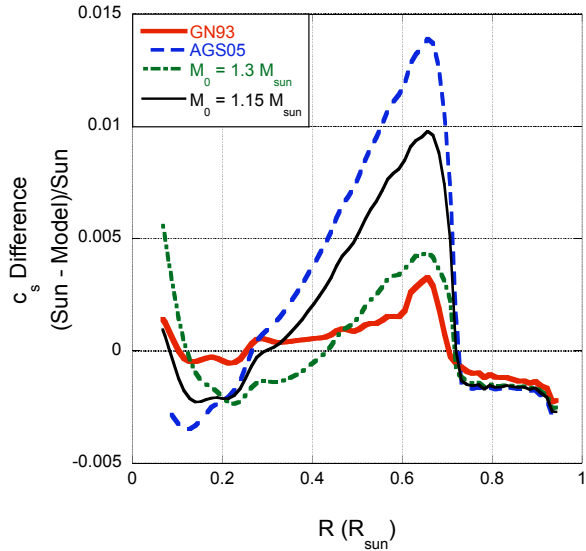}}
 \caption{\footnotesize  Inferred minus calculated sound speed differences for calibrated standard one solar-mass models using the GN93 and AGS05 abundances and for models with AGS05 abundances and initial mass 1.3 and 1.15 M$_{\odot}$ including early mass loss.}
 \label{fig:c_massloss}
\end{figure}

\begin{figure}[t!]
% \resizebox{\hsize}{!}{\includegraphics{O-Cstvsmlmodels.eps}}
 \resizebox{\hsize}{!}{\includegraphics{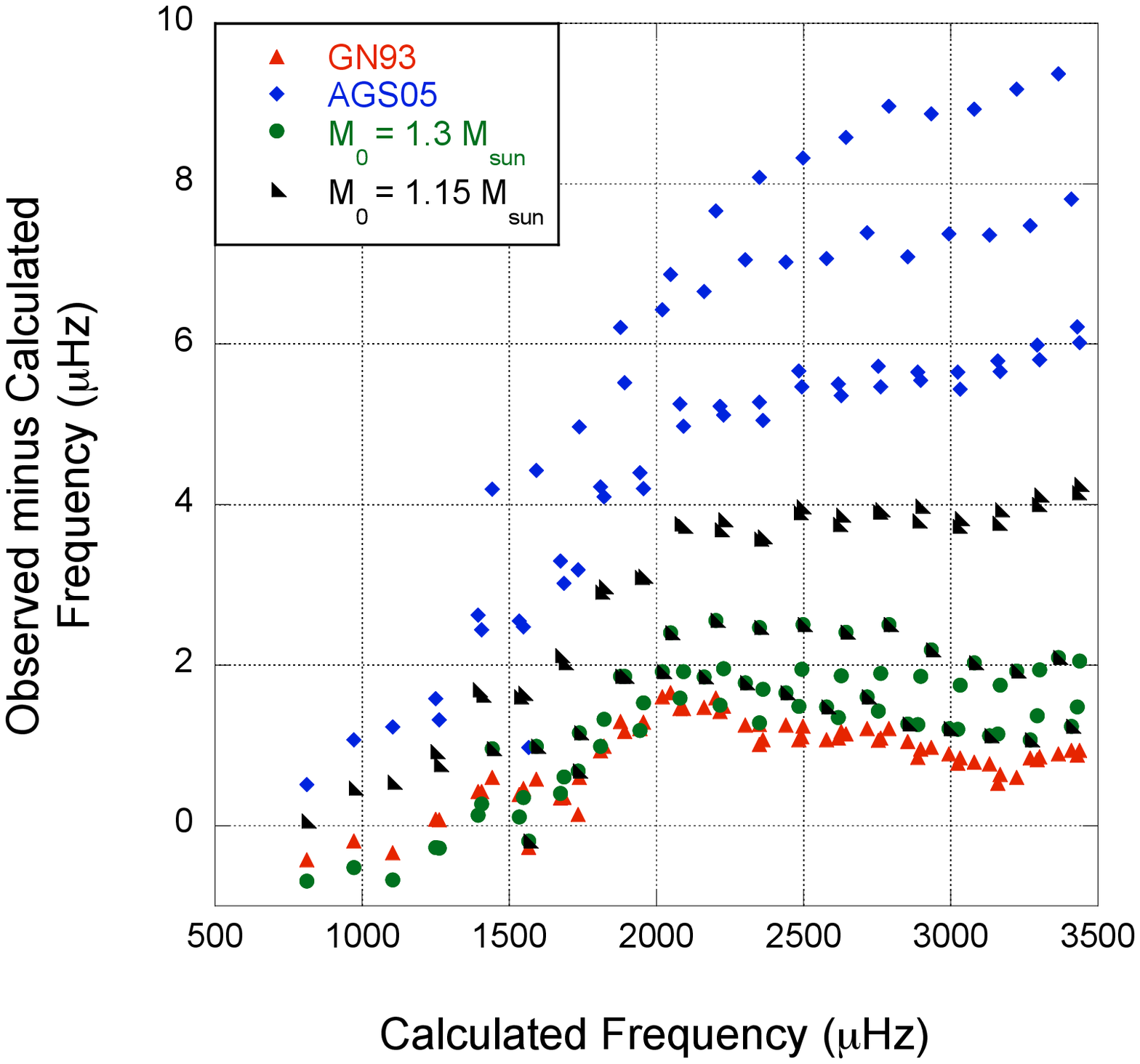}}
 \caption{\footnotesize  Observed minus calculated versus calculated frequencies for calibrated standard one solar mass models using the GN93 and AGS05  abundances, and for models with AGS05 abundances with initial mass 1.3 and 1.15 M$_{\odot}$ including early mass loss. Frequencies compared are for modes of angular degrees $\ell$ = 0, 2, 10, and 20.  The data are from \citet{Chaplin_2007}, \citet{ST96}, and \citet{Garcia_2001}. The calculated frequencies were computed using the \citet{Pesnell_1990} non-adiabatic stellar pulsation code.}
 \label{fig:O-C_massloss}
\end{figure} 

\begin{figure}[t!]
% \resizebox{\hsize}{!}{\includegraphics{Litempvs.time.eps}}
 \resizebox{\hsize}{!}{\includegraphics{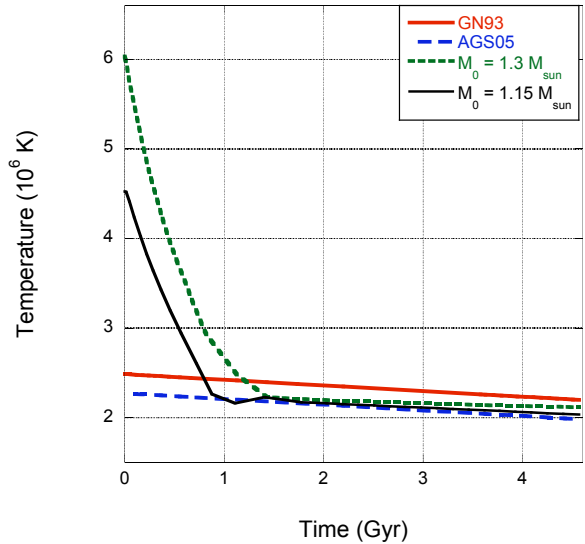}}
 \caption{\footnotesize  Temperature experienced by the present-day solar surface layer as a function of time for the mass-losing and standard models.  For the mass-losing phase, the lithium-destroying temperatures are attained because the layer that is now at the surface once resided deeper inside the sun. In the post-mass-loss phase, the relevant temperatures are attained by envelope convection which mixes surface layers downward, exposing the surface material to the temperature at the CZ base.  2.8 million K is the temperature required for relatively rapid Li destruction.}
 \label{fig:temp_massloss}
\end{figure}

Table \ref{table:ZYR} summarizes the initial Y and mixing-length parameter needed to calibrate each model and the final CZ Y and CZ base radius. The seismically-inferred CZ Y abundance and CZ base radius are 0.248 $\pm$ 0.003 and 0.713 $\pm$ 0.001 R$_{\odot}$, respectively \citep{BA04a}. Figure \ref{fig:lum_massloss} shows the luminosity versus time for these models, as well as for two constant, one solar-mass calibrated models. Figure \ref{fig:c_massloss} shows the inferred minus calculated sound speed for these models.  For the models with the AGS05 abundances, the sound speed agreement is considerably improved by including early mass loss.  For the model with initial mass 1.3 M$_{\odot}$ sound-speed agreement is almost restored near the CZ base, but the agreement is not as good in the more H-depleted core.   Unfortunately, while the model with initial mass 1.15 M$_{\odot}$ has a little better sound speed agreement in the central 0.1 R$_{\odot}$, the improvement is not as pronounced for the region below the CZ. Figure \ref{fig:O-C_massloss} shows the observed minus calculated versus calculated nonadiabatic frequencies for modes of angular degrees $\ell$ = 0, 2, 10, and 20 that propagate into the solar interior below the convection zone. Including mass loss improves agreement, but models with old abundances and no mass loss still give the best agreement.

The mass-losing models described here would destroy all of the observed surface lithium.  Lithium is destroyed relatively rapidly in the solar interior at temperatures  $\ge$ 2.8 million K. For standard models, on the main sequence the surface layers are never mixed to high enough temperatures to deplete Li by the observed factor of 150 from the initial solar system abundance \citep{AGSS09}, and additional mixing mechanisms must be invoked.  However, with mass loss, layers that are now at the surface were initially in the interior at temperatures high enough to quickly destroy Li. Figure \ref{fig:temp_massloss} shows the temperature experienced by the surface layer throughout the evolution of each model. During the mass-losing phase, the high temperatures that depleted the lithium in the current surface layer were experienced when the material that is at the surface of the now 1 M$_\odot$ sun was deeper, before the previous surface layers were lost. After the mass-loss phase, the temperature that affects the surface layer is that of the CZ base, since the material currently at the surface is continually mixed through the CZ.

The mass-losing models also produce more ${}^3\rm{He}$ at the surface as the now-surface layers were once processed at higher interior temperatures where ${}^3\rm{He}$ builds up to higher equilibrium values.  For the 1.3 M$_{\odot}$ initial mass model, the surface ${}^3\rm{He}$ mass fraction is enhanced from its initial value of 5.0 $\times10^{-5}$ to 9.0 $\times10^{-5}$, while for the 1.15 M$_{\odot}$ initial mass model, the surface ${}^3\rm{He}$ mass fraction is only slightly enhanced from its initial value of 5.0 $\times10^{-5}$ to 5.1 $\times10^{-5}$. The final ${}^3\rm{He}/{}^4\rm{He}$ abundance ratios for the 1.3 and 1.15 M$_{\odot}$ initial mass models are 3.9 $\times10^{-4}$ and 2.2 $\times10^{-4}$, respectively, enhanced from an initial value of 2.0 $\times10^{-4}$.

\section{Accretion of low-Z material}
\label{sect:accret}

\subsection{Motivation and method}
\label{sect:acc_meth}

As a way to keep the solar interior more like models obtained using higher abundances, \citet[][hereafter GWC04]{GWC04} and GWC05 proposed accretion of material depleted in heavier elements early in the sun's lifetime. In this scenario, the pre-main sequence sun would have $\sim$98$\%$ of its present mass and a higher Z with a mixture similar to the GN93 or GS98 abundances. After the sun begins core hydrogen burning and is no longer fully convective, the remaining $\sim$2$\%$ of material accreted would have lower Z providing a convection-zone abundance similar to the current photospheric abundances of AGS05 or AGSS09. Possible justifications for this scenario are discussed by \citet{Nordlund_2009} and \citet{Melendez_2009}. One plausible explanation is that planet formation removes some high-Z elements from the solar nebula, leaving lower-Z material to be accreted after the sun is no longer fully convective.

\citet{Winnick_2002} explored the accretion of metal-rich material in models with the older GS98 abundances. They show that some solar models with enhanced metallicity in the convection zone might be viable as small perturbations to the standard GS98 model. \citet{Haxton_2008} discuss the question of accretion of metal-depleted gas onto the sun as a motivation for future experiments to measure CN-cycle neutrinos. The flux of these neutrinos should depend nearly linearly on the initial core abundance of C and N. A successful measurement of the CN-cycle solar neutrino flux would therefore place constraints on possible accretion by determining the metallicity of the solar core.

To test the possibility of low-Z accretion in the early sun, a model is evolved starting with $Z=0.0197$ on the zero-age main sequence, and material is progressively added to reduce the CZ Z by $0.001$ in each of six steps of about six million years. After each step, the model is given time to equilibrate to a new shallower CZ depth, leaving behind the higher-Z composition gradient \citep{Guzik_2006}. The final accretion episode leaves the CZ with $Z=0.0137$. After 36 million years of low-Z accretion, the model is evolved normally (including diffusive settling) and calibrated as usual to the observed luminosity, radius, and AGS05 $Z/X$ value.

\subsection{Results}
\label{sect:acc_results}

\begin{figure}[t!]
% \resizebox{\hsize}{!}{\includegraphics{Zaccretionbwplot.eps}}
% \resizebox{\hsize}{!}{\includegraphics{Zaccretion.plot.eps}}
 \resizebox{\hsize}{!}{\includegraphics{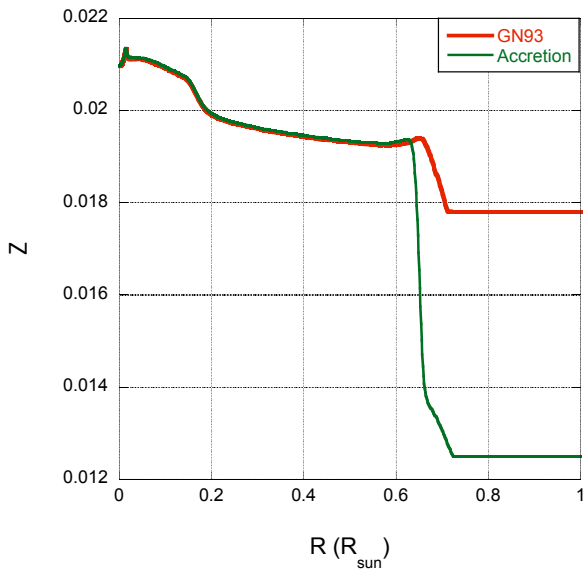}}
 \caption{\footnotesize Heavy element abundance profiles for the GN93 mixture model and the low-Z accretion model.}
 \label{fig:Z_accret}
\end{figure}

\begin{figure}[t!]
% \resizebox{\hsize}{!}{\includegraphics{Yaccretion.plot.eps}}
 \resizebox{\hsize}{!}{\includegraphics{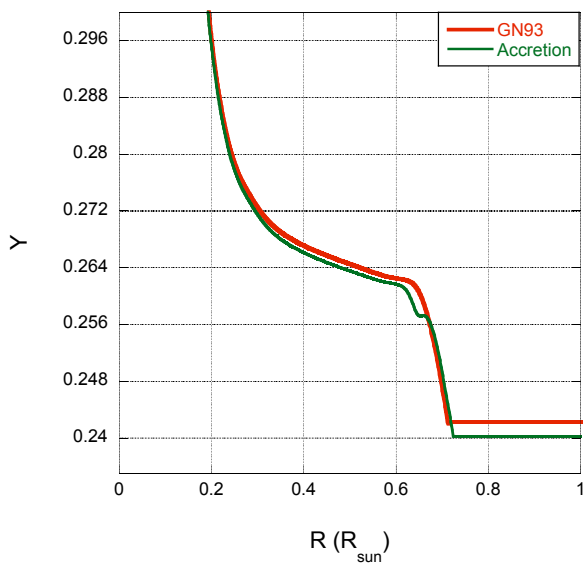}}
 \caption{\footnotesize Helium abundance profiles for the GN93 mixture model and the low-Z accretion model.}
 \label{fig:Z_accret_Y}
\end{figure}

Figure \ref{fig:Z_accret} shows the heavy element abundance throughout the sun in the accretion model, and Figure \ref{fig:Z_accret_Y} shows the Y abundance. As intended, the abundance in the interior is similar to the GN93 model while the CZ abundance matches the new, lower Z from photospheric observations. 

\begin{figure}[t!]
% \resizebox{\hsize}{!}{\includegraphics[width=150mm]{accretcombinedplot.eps}}
 \resizebox{\hsize}{!}{\includegraphics[width=150mm]{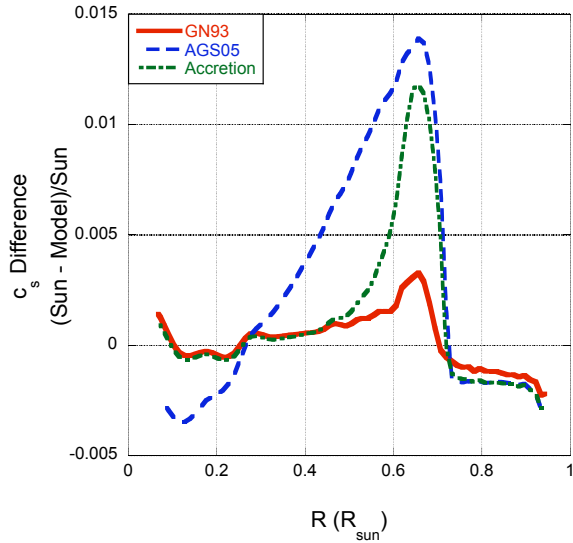}}
 \caption{\footnotesize Relative difference between inferred and calculated sound  speeds for models with the GN93 and AGS05 abundances and with low-Z accretion.}
 \label{fig:c_accret}
\end{figure}

\begin{figure}[t!]
% \resizebox{\hsize}{!}{\includegraphics{ominc+accret.plot.eps}}
 \resizebox{\hsize}{!}{\includegraphics{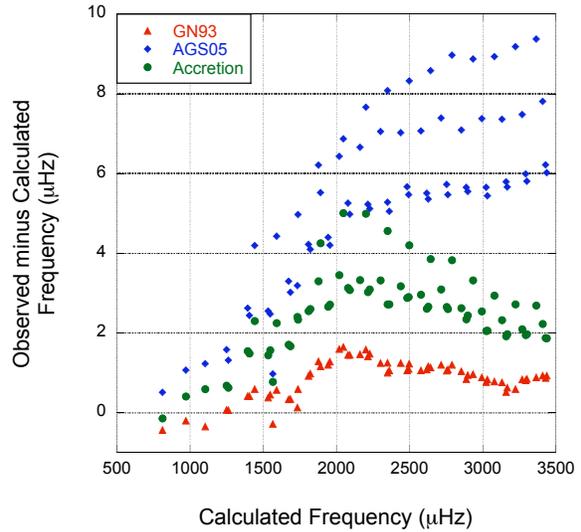}}
 \caption{\footnotesize  Observed minus calculated frequency versus calculated ferquency of GN93, AGS05, and accretion models for degree $\ell$=0, 2, 10, and 20 modes. The calculated frequencies were computed using the \citet{Pesnell_1990} non-adiabatic stellar pulsation code. The data are from \citet{Chaplin_2007}, \citet{ST96}, and \citet{Garcia_2001}. The accretion model shows improved, though not acceptable, O-C agreement.}
 \label{fig:O-C_accret}
\end{figure}

Figure \ref{fig:c_accret} shows the relative sound-speed differences for models with the GN93 mixture, the AGS05 mixture, and low-Z accretion. The accretion model shows improvement in sound-speed agreement in the interior where Z is similar to the GN93 mixture. However, discrepancy remains near the CZ base. Compared to the AGS05 model, the accretion model has a less shallow CZ base radius of 0.7235 R$_{\odot}$ and a nearly acceptable CZ Y abundance of $0.2407$. Figure \ref{fig:O-C_accret} shows the observed minus calculated versus calculated non-adiabatic frequencies for modes of angular degrees $\ell$ = 0, 2, 10, and 20 that propagate into the solar interior below the CZ. Including accretion in a model using the new abundances improves agreement with this frequency data.

\citet{CVR07} also approximated an accretion model by instantaneously decreasing the Z abundance gradient in the CZ in an early main-sequence model (age 74 My). They do not find an improvement in the CZ depth, as we did for our model, but find about the same CZ Y abundance, $0.240$, and improved sound-speed agreement below 0.5 R$_\odot$.

\section{Convective overshoot}
\label{sect:overshoot}

\subsection{Motivation and method}
\label{sect:over_meth}

It is possible that the CZ depth predicted using standard mixing-length theory is too shallow and convective motions extend the nearly adiabatically stratified part of the CZ to the depth inferred seismically. \citet{Rempel_2004} presents a semi-analytical model of overshoot. This approach facilitates the understanding of the relation between numerical simulations and classical overshoot theories in terms of physical parameters. \citet{Montalban_2006} developed solar models that include convective overshoot. By adopting an overshoot parameter of the order of  0.15 times the pressure scale height and increasing the opacity by $\sim$7$\%$ (within the uncertainty limits of the abundances), they were able to reproduce the seismically inferred CZ base and Y$_{CZ}$. However, large sound-speed discrepancies remain in the radiative region of their model.

To explore the possibility of convective overshoot, we evolve models with AGS05 abundances but extend the CZ that follows the adiabatic gradient to a depth that optimizes agreement with the sound speed inversions. We hoped that a deeper CZ would also inhibit diffusion and keep the CZ Y abundance higher.

\subsection{Results}
\label{sect:over_results}

\begin{figure}[t!]
% \resizebox{\hsize}{!}{\includegraphics{soundovershoot.plot.eps}}
 \resizebox{\hsize}{!}{\includegraphics{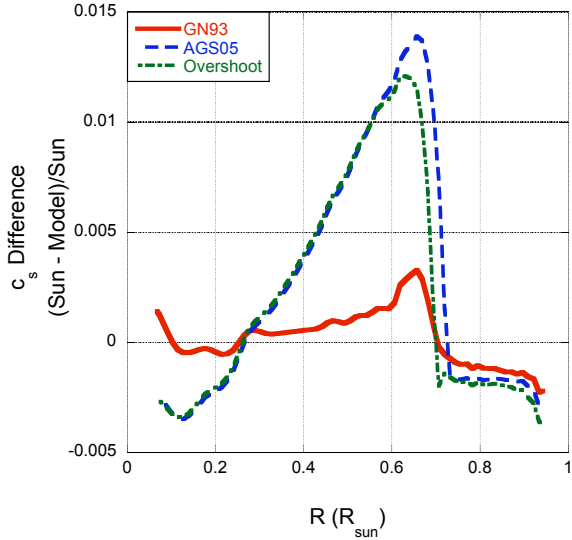}}
 \caption{\footnotesize Relative difference between inferred and calculated sound  speeds for models with the GN93 and AGS05 abundances and the first convective overshoot model that extends the adiabatically stratified layer to 0.704 R$_\odot$.}
 \label{fig:c_over1}
\end{figure}

\begin{figure}[t!]
% \resizebox{\hsize}{!}{\includegraphics{soundovertoomuch.plot.eps}}
 \resizebox{\hsize}{!}{\includegraphics{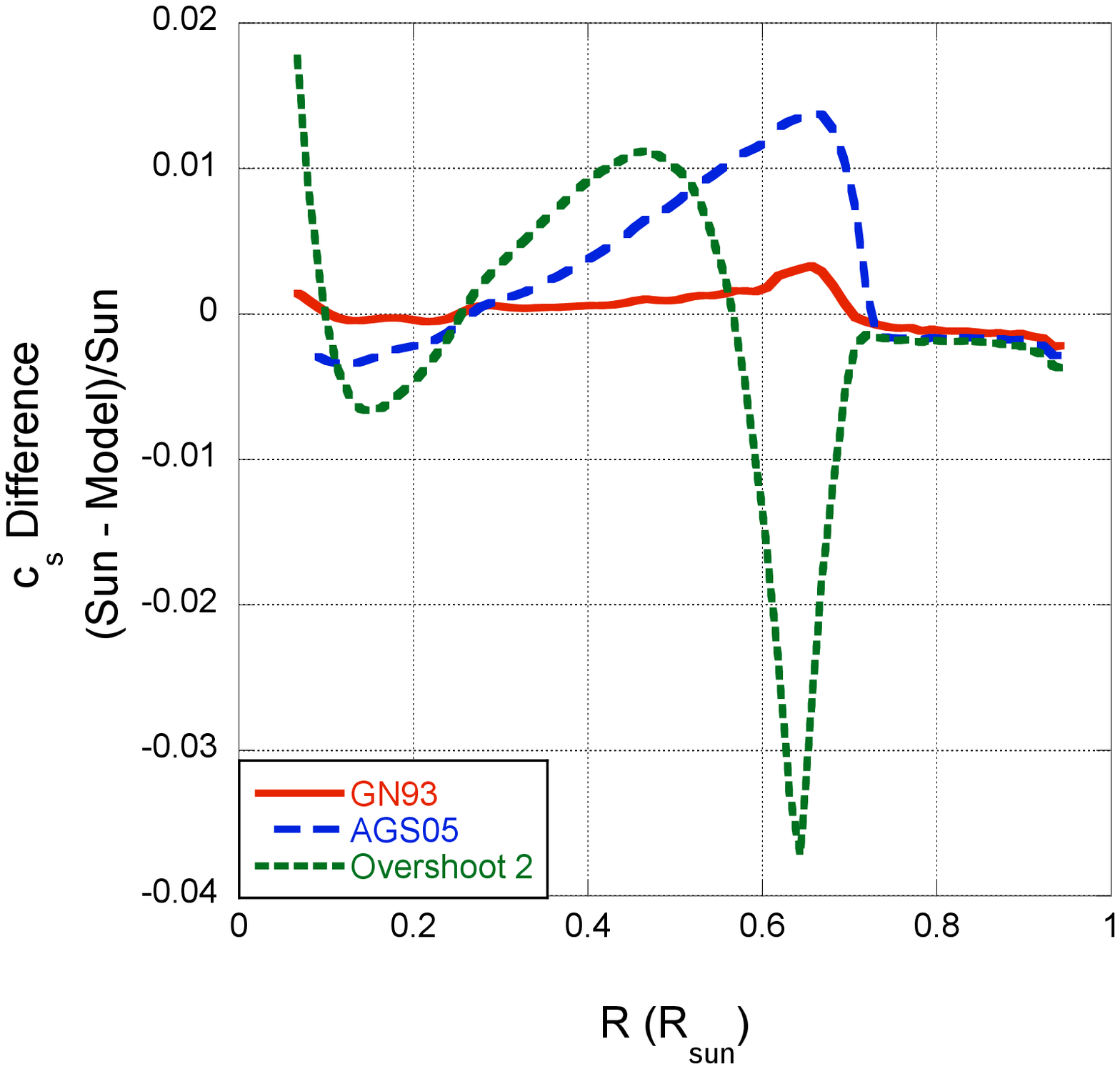}}
 \caption{\footnotesize Relative difference between inferred and calculated sound speeds for models with the GN93 and AGS05 abundances and the second convective overshoot model that extends the adiabatically stratified layer to 0.64 R$_\odot$.}
 \label{fig:c_over2}
\end{figure}

\begin{figure}[t!]
% \resizebox{\hsize}{!}{\includegraphics{O-Covershoot.plot.eps}}
 \resizebox{\hsize}{!}{\includegraphics{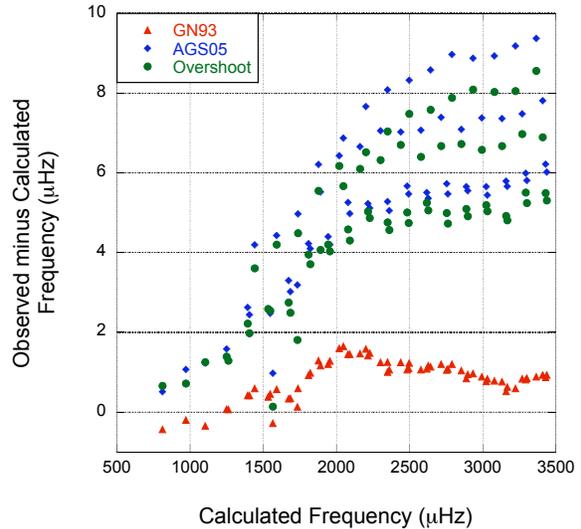}}
 \caption{\footnotesize Observed minus calculated frequency versus calculated frequency for models with the GN93 and AGS05 abundances and for the overshoot model for degree $\ell$=0, 2, 10, and 20 modes. The calculated frequencies were computed using the \citet{Pesnell_1990} non-adiabatic stellar pulsation code, and the data are from \citet{Chaplin_1998}, \citet{ST96}, and \citet{Garcia_2001}.}
 \label{fig:O-C_over}
\end{figure}

Figure \ref{fig:c_over1} shows the relative sound-speed differences for models with the GN93 mixture, the AGS05 mixture, and the first convective overshooting model. For the first overshoot model, the CZ depth is 0.704 R$_{\odot}$, deeper than inferred seismically. The sound speed agreement is improved only within the CZ but not much below it. The deeper CZ does not inhibit Y diffusion as we had hoped. In the second model, we extend the CZ even deeper, to 0.64 R$_{\odot}$. Figure \ref{fig:c_over2} shows the relative sound-speed difference for the second convective overshooting model and standard models with the GN93 and AGS05 mixtures. The sound speed gradient at the base of this adiabatically stratified CZ clearly does not agree with the seismically inferred one, and sound speed agreement in the central 0.2 R$_{\odot}$ is much worse than in any of the other models. We therefore omit the second overshoot model from further analysis. Figure \ref{fig:O-C_over} shows the observed minus calculated frequency for the first overshoot model compared to the GN93 and AGS05 models. The AGS05 model and the overshoot model where the core Z is low do not agree as well with the data as the GN93 model does. It appears that overshooting alone is not a solution to this problem.

\section{Summary of mitigation attempts}
\label{sect:summary}

%\begin{table}
%\caption{CZ Y, CZ base radius, and photosphere Z/X}
%\label{table:ZYR}
%\begin{center}
%\begin{tabular}{llll}
%hline
%Model                  &  Y$_{CZ}$  &  R$_{CZ Base}$ & Z/X  \\
%\hline
%GN93                   & $0.2418 $  & $0.7133$  &  $0.0244$ \\
%AGS05                  & $0.2273 $  & $0.7306$  &  $0.0163$ \\
%Mass loss 1            & $0.2324 $  & $0.7195$  &  $0.0173$ \\
%Mass loss 2            & $0.2297 $  & $0.7231$  &  $0.0168$ \\
%Accretion              & $0.2407 $  & $0.7235$  &  $0.0170$ \\
%Overshoot              & $0.2292 $  & $0.7038$  &  $0.0164$ \\
%Seismic                & $0.2485 \pm 0.0035$  & $0.713 \pm 0.001$   &  \\
%\hline
%\end{tabular}
%\tablenotetext{a}{Mass loss models 1 and 2 have initial masses 1.3 M$_\odot$ and 1.15 M$_\odot$, respectively.}
%\tablenotetext{b}{Seismically inferred values from \citet{BA04a}.}
%\end{center}
%\end{table}

\begin{table*}
\caption{Initial mass and surface abundances, mixing length parameter, and final abundances and CZ base for our solar models.}
\label{table:ZYR}
\begin{center}
\begin{tabular}{llllllllll} 
\hline
Model & GN93 & AGS05 & ML 1 & ML 2  & Accretion & Overshoot  & C 1 & C 2 & C 3 \\
\hline
M$_{\rm{o}}$/M$_\odot$& 1.00 & 1.00 & 1.30 & 1.15 & 0.98 & 1.00 & 1.00 & 1.00 & 1.10 \\
Y$_{\rm{o}}$ & 0.26930 & 0.25700 & 0.24659 & 0.25279 & 0.26927 & 0.25699 & 0.26370 & 0.27780 & 0.27530 \\
Z$_{\rm{o}}$  & 0.01970 & 0.01350 & 0.01351 & 0.01351 & 0.01973 & 0.01351 & 0.01740 & 0.01700 & 0.01700 \\
$\alpha$& 2.0379 & 2.0004 & 2.0571 & 2.0104 & 1.8958 & 1.9962 & 1.9918 & 2.0635 & 2.0652  \\
Z/X      & 0.0240 & 0.0163 & 0.0178 & 0.0171 & 0.0162 & 0.0164 & 0.0209 & 0.0208 & 0.0219  \\
Y$_{\rm{CZ}}$& 0.2412 & 0.2273 & 0.2388 & 0.2349 & 0.2402 & 0.2292 & 0.2349 & 0.2473 & 0.2551  \\
R$_{\rm{CZB}}$ /R$_\odot$& 0.7125 & 0.7294 & 0.7217 & 0.7264 & 0.7241 & 0.7038 & 0.7186 & 0.7190 & 0.7181 \\
\hline
\end{tabular}
\tablenotetext{a}{Seismically inferred values from \citet{BA04a}: Y$_{\rm{CZ}}$ =  $0.2485 \pm 0.0035$, R$_{\rm{CZB}}$ /R$_\odot$ =  $0.713 \pm 0.001$.}
\tablenotetext{b}{Models ML 1 and ML 2 are the AGS05 models including mass loss. Models C1, C2, and C3 are the CO$^5$BOLD models with GN93 opacities, AGS05 opacities, and AGS05 opacities including mass loss.}
\end{center}
\end{table*}

\begin{figure}[t!]
 \resizebox{\hsize}{!}{\includegraphics{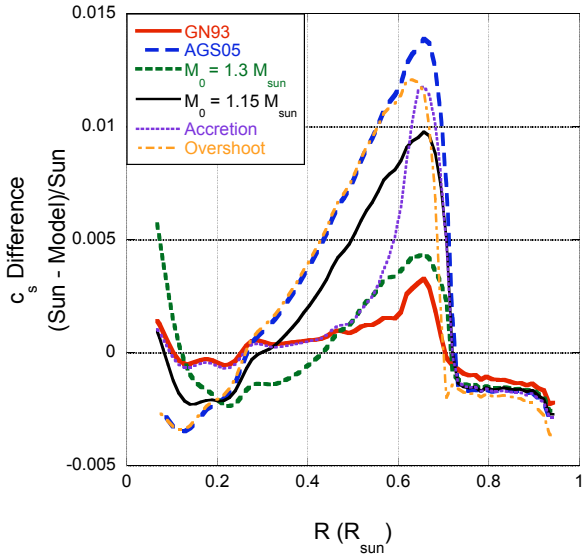}}
 \caption{\footnotesize  Relative difference between inferred and calculated sound speeds for models with the GN93 and AGS05 abundances and models with mass-loss, accretion, and convective overshoot.} 
 \label{fig:c_all}
\end{figure}

\begin{figure}[t!]
 \resizebox{\hsize}{!}{\includegraphics{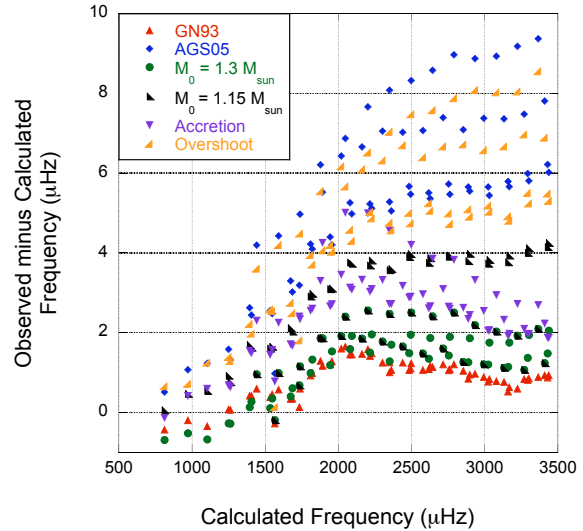}}
 \caption{\footnotesize   Observed minus calculated frequency versus calculated frequency for degree $\ell$=0, 2, 10, and 20 modes in GN93 and AGS05 models and in models with mass-loss, accretion, and convective overshoot. The calculated frequencies were computed using the \citet{Pesnell_1990} non-adiabatic stellar pulsation code. The data are from \citet{Chaplin_2007}, \citet{ST96}, and \citet{Garcia_2001}.}
 \label{fig:O-C_all}
\end{figure}

\begin{figure}[t!]
 \resizebox{\hsize}{!}{\includegraphics{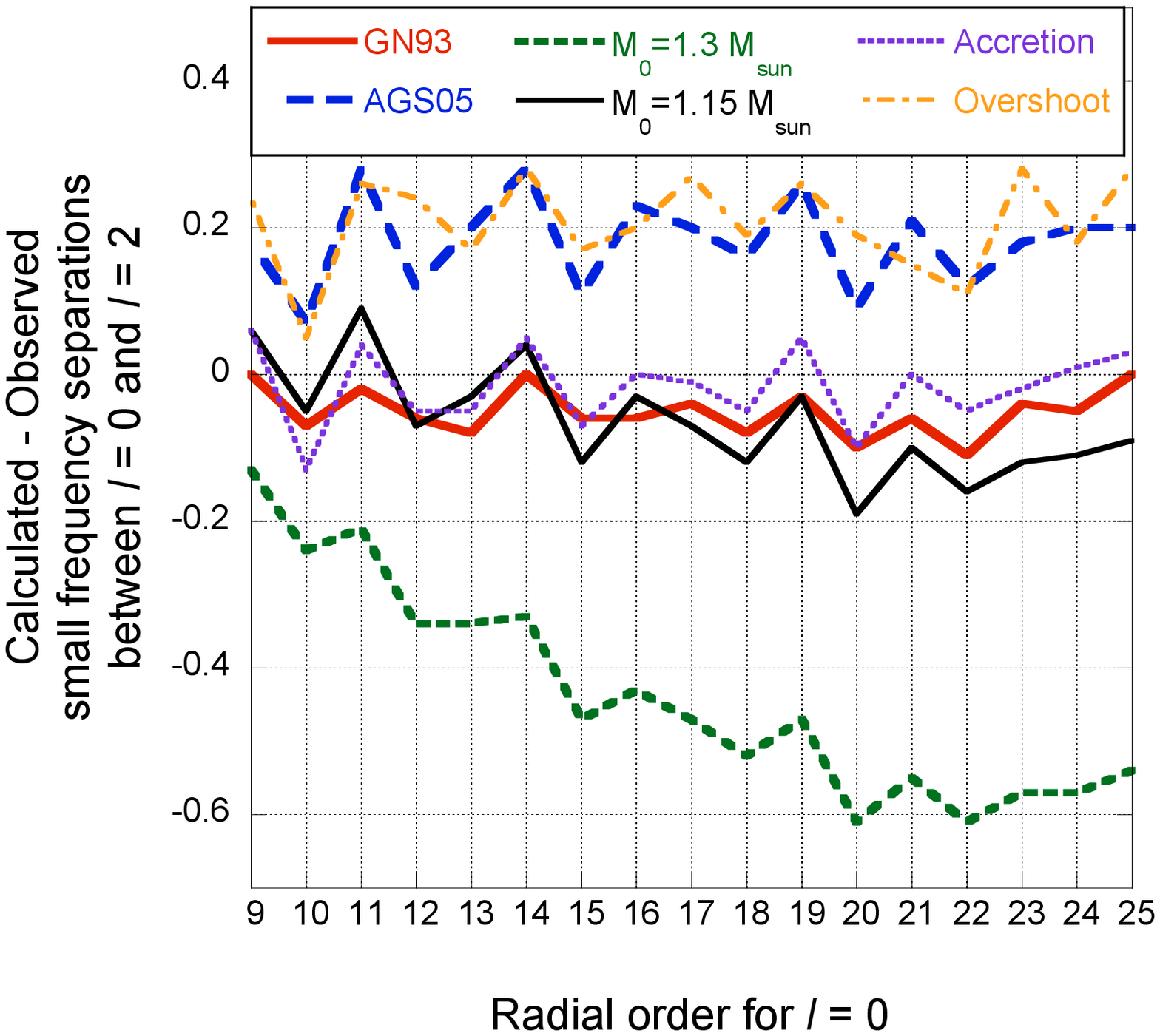}}
% \resizebox{\hsize}{!}{\includegraphics{small_separations_2007.eps}}
 \caption{\footnotesize  Difference between calculated and observed small separations for $\ell$=0 and 2 modes for the Guzik et al. models. The data are from \citet{Chaplin_2007}.}
 \label{fig:small_sep_all}
\end{figure}

Table \ref{table:ZYR} summarizes the CZ Y, CZ base radius, and photosphere Z/X for the models examined here. Figure \ref{fig:c_all} shows the relative sound-speed differences of our models with the GN93 mixture, the AGS05 mixture, and models with mass-loss, accretion, and convective overshoot. The observed minus calculated frequencies of these models are shown in Figure \ref{fig:O-C_all}. As seen in Table \ref{table:ZYR} and Figures \ref{fig:c_all} and \ref{fig:O-C_all}, the AGS05 model and overshoot model do not agree well with the data. The GN93 model, the 1.3 M$_\odot$ mass loss model, and the accretion model show better agreement, though no model matches the data perfectly. 
 
Figure \ref{fig:small_sep_all} shows the small frequency separation differences of the Guzik et al. models minus the solar-cycle corrected frequency differences from the BiSON group \citep{Chaplin_2007} for $\ell$=0 and 2 modes, which are sensitive to the structure of the core. This plot illustrates that including low-Z accretion in the model retains the core structure of the GN93 model. The overshoot model and AGS05 model do not agree as well with the data as the models with higher core Z. The mass losing model with the best sound speed and observed minus calculated frequency agreement (M$_0$ = 1.3 M$_{\odot}$) shows worse agreement than any of the other models. However, mass loss does change the value of the small frequency separation in the right direction to correct for the discrepancy seen with the new abundances. Perhaps the over-compensation indicates that this model has too much mass loss; a model with a smaller initial mass would reduce the disagreement, as seen with the M$_0$ = 1.15 M$_{\odot}$ model.

Mass-losing models can improve seismic agreement for the new abundances, but they do not fully restore agreement. In addition, the destruction of too much Li and the production of too much surface ${}^3\rm{He}$ make the two models considered here unlikely. A smaller amount of mass loss that leads to destruction of some, but not all, of the initial Li could provide a plausible partial mitigation of the solar abundance problem.

The accretion model allows for a solar interior that is similar to models developed with the higher abundances and therefore agrees nicely with seismic inferences in the central 0.5 R$_{\odot}$ of the sun. The model still shows poor agreement near the CZ base. The very steep Z abundance gradient developed at the CZ base seen in Figure \ref{fig:Z_accret} \citep[see][]{Guzik_2006} might have a detectable signature in the seismic frequencies. \citet{Basu_1997} finds that inversions appear to rule out such steep composition gradients at the CZ base.
    
The convective overshoot model with a CZ depth of 0.704 R$_{\odot}$ improves sound speed agreement slightly within the CZ but not much below it. Extending the CZ depth to 0.64 R$_{\odot}$ results in even worse agreement. In addition, Y diffusion is not inhibited by the deeper CZ, as we had hoped.

\section{CO$^5$BOLD abundances}
\label{sect:COBOLD}

At the suggestion of both the referee and a colleague P. Bonifacio, we 
also include here a preliminary exploration of solar models using the 
CO$^5$BOLD abundances with a Z/X of 0.0209 and Z of 0.0154, intermediate 
between the AGS05 and AGSS09 abundances, but lower than the GN93 
abundances.

Because the abundances of only 12 elements have been re-evaluated at 
this time by the CO$^5$BOLD group \citep{Ludwig_2009}, it is a little 
premature to create new opacity tables for the CO$^5$BOLD mixture, which 
can, in principle be done using the Lawrence Livermore OPAL web request 
at http://opalopacity.llnl.gov. In addition, we do not have 
low-temperature opacities for a mixture representative of the CO$^5$BOLD 
abundances available to us.  Therefore, here we decided to calibrate 
standard models to the Z/X of CO$^5$BOLD using opacity tables based on the 
GN93 and AGS05 mixtures.  We observe that the O/Fe (oxygen to iron) mass 
fraction ratio of CO$^5$BOLD is 4.98, intermediate between the O/Fe mass 
ratio of 6.56 for the GN93 mixture and 4.65 for the AGS05 mixture, so 
our two standard models should bracket results that use the 
CO$^5$BOLD mixture in the opacity tables.  We did update abundances to the 
CO$^5$BOLD values for our in-line equation of state calculation and for 
tracking the diffusion of the major elements.
 
\begin{figure}[t!]
% \resizebox{\hsize}{!}{\includegraphics{sound_Caffau.eps}}   
 \resizebox{\hsize}{!}{\includegraphics{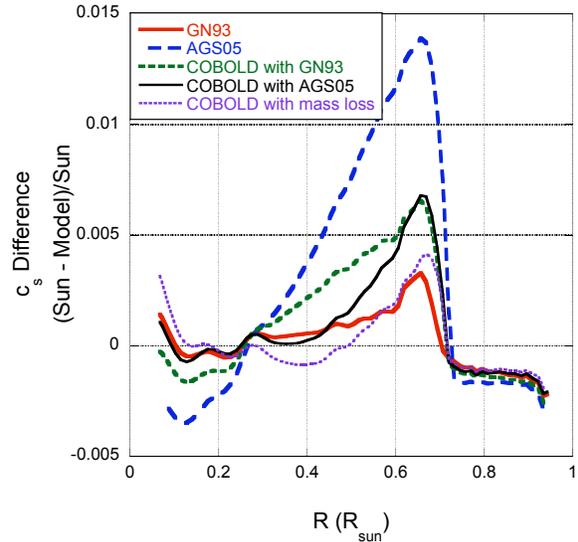}}
 \caption{\footnotesize Relative difference between inferred and calculated sound speeds for models using the CO$^5$BOLD abundances created with either the GN93 or AGS05 opacities and a model created using the CO$^5$BOLD abundances with the AGS05 opacities and including mass loss.}
 \label{fig:c_Caf}
\end{figure}

\begin{figure}[t!]
% \resizebox{\hsize}{!}{\includegraphics{smallsep_Caffau.eps}}
 \resizebox{\hsize}{!}{\includegraphics{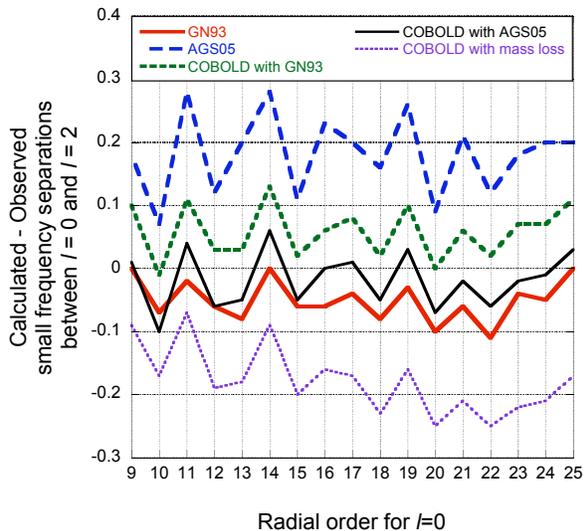}}
 \caption{\footnotesize Difference between calculated and observed small separations for $\ell$=0 and 2 modes for the Guzik et al. models using the CO$^5$BOLD abundances with either the GN93 or AGS05 opacities and with the AGS05 opacities and mass loss. The data are from \citet{Chaplin_2007}.}
 \label{fig:smallsep_Caf}
\end{figure}

\begin{figure}[t!]
 \resizebox{\hsize}{!}{\includegraphics{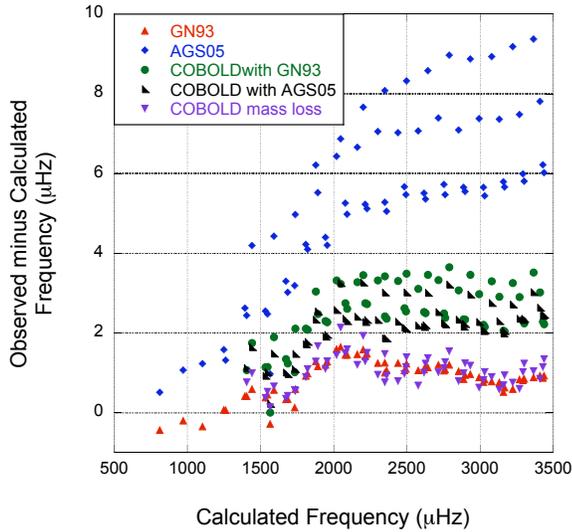}}
 \caption{\footnotesize   Observed minus calculated frequency versus calculated frequency for degree $\ell$=0, 2, 10, and 20 modes in  models using the CO$^5$BOLD abundances with either the GN93 or AGS05 opacities and with the AGS05 opacities and mass loss. The calculated frequencies were computed using the \citet{Pesnell_1990} non-adiabatic stellar pulsation code. The data are from \citet{Chaplin_2007}, \citet{ST96}, and \citet{Garcia_2001}.}
 \label{fig:O-C_Caf}
\end{figure}

The initial Y, Z, and mixing length to pressure-scale-height ratios 
needed to calibrate to the CO$^5$BOLD Z/X using either opacity set are 
listed in Table \ref{table:ZYR}; Figures \ref{fig:c_Caf}, \ref{fig:smallsep_Caf}, and \ref{fig:O-C_Caf} show the results for sound-speed 
differences, small separations between calculated $\ell$=0 and $\ell$=2 modes, and observed minus calculated frequencies for $\ell$ = 0, 2, 10, and 20. The sound speed discrepancy is reduced to only about 0.6$\%$ at the 
convection zone base for the CO$^5$BOLD abundances, compared to 1.4$\%$ for 
the AGS05 abundances and 0.4$\%$ for the GN93 abundances. The results for 
either opacity set are identical above 0.6 R$_\odot$, but differ below this 
where oxygen is the main opacity contributor, and in the core where iron 
is the main opacity contributor, as expected.  The model using the AGS05 
opacity mixture requires a higher Y abundance to compensate for the 
relatively higher Fe opacity contribution in the core.  The small 
separations and observed minus calculated frequencies are slightly higher than found for a model calibrated to the GN93 abundances on average, but not as 
high as for a model calibrated to the AGS05 Z/X.
 
As also surmised by the referee, since the CO$^5$BOLD sound speed 
differences are closer to observed, improvement could be obtained with a 
smaller amount of mass loss.  Here we have calculated an additional mass-loss model with initial mass 1.1 M$_\odot$ and an initial mass-loss rate of 2.25 x 10$^{-10}$ M$_\odot$/yr, exponentially decaying with e-folding time 0.45 Gyr.  This model was calibrated to the CO$^5$BOLD Z/X using the AGS05 
opacities that have O/Fe abundance closer to that of the CO$^5$BOLD O/Fe.  
Previous work \citep[e.g.][]{SF92,GC95,SB03} indiciated that an initial mass of 1.1 M$_\odot$ or less and a relatively short mass loss phase (less than 0.2-0.5 Gyr) could deplete the lithium to the present-day observed values from initial solar-system abundance without completely destroying the lithium or building up too much $^3$He.  The sound speed agreement (Figure \ref{fig:c_Caf}) shows considerable improvement with this smaller amount of mass loss; however, the agreement for the solar 
core is not as good as for the non-mass losing models, as can be seen 
more clearly in the small separations (Figure \ref{fig:smallsep_Caf}). This mass-losing model restores the level of agreement attained with the GN93 abundances for the observed minus calculated frequencies (Figure \ref{fig:O-C_Caf}).  

\begin{figure}[t!]
 \resizebox{\hsize}{!}{\includegraphics{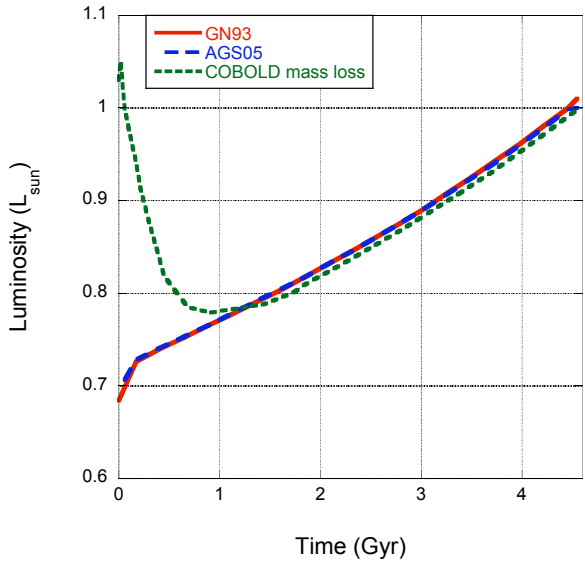}}
 \caption{\footnotesize Luminosity versus time for standard one solar-mass models using the GN93 and AGS05 abundances and for a mass-losing model using the CO$^5$BOLD abundances with the AGS05 opacities with initial mass 1.1 M$_{\odot}$.  Mass-loss rates are exponentially decaying with e-folding time 0.45 Gyr.}
 \label{fig:lum_CafML}
\end{figure}

\begin{figure}[t!]
 \resizebox{\hsize}{!}{\includegraphics{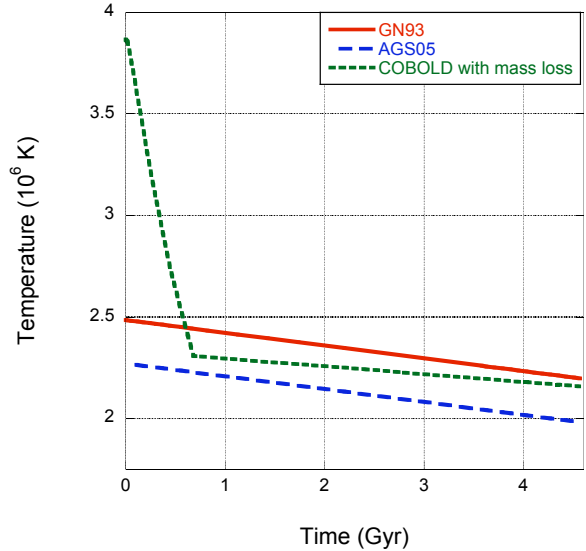}}
 \caption{\footnotesize  Temperature experienced by the present-day solar surface layer as a function of time for the standard models and the mass-losing model with CO$^5$BOLD abundances and AGS05 opacities.  For the mass-losing phase, the lithium-destroying temperatures are attained because the layer that is now at the surface once resided deeper inside the sun. In the post-mass-loss phase, the relevant temperatures are attained by envelope convection which mixes surface layers downward, exposing the surface material to the temperature at the CZ base.  2.8 million K is the temperature required for relatively rapid Li destruction.}
 \label{fig:temp_CafML}
\end{figure}

Figure \ref{fig:lum_CafML} shows the luminosity versus time for this model, as well as for the standard solar models evolved with GN93 and AGS05 abundances. Figure \ref{fig:temp_CafML} shows the effective temperature experienced by the surface layer throughout the evolution of the mass-loss model compared to that experienced by the standard models. During the mass-losing phase, the high temperatures that depleted the lithium in the current surface layer were experienced when the material that is at the surface of the now 1 M$_\odot$ sun was deeper, before the previous surface layers were lost. After the mass-loss phase, the temperature that affects the surface layer is that of the CZ base, since the material currently at the surface is continually mixed through the CZ. We see that including mass loss in the CO$^5$BOLD model exposes the surface layers to high enough temperatures to deplete Li early in the evolution (2.8 million K is the temperature required for relatively rapid Li destruction).
It is not clear that the advantages of mass loss (e.g. Li depletion and better sound speed agreement in the outer 80$\%$ of the solar radius) can be retained while at the same time not creating a discrepancy in the inner 20$\%$.

\section{Conclusions and future work}
\label{sect:conc}

In spite of the seismic evidence in favor of the old abundances, the new abundances cannot be easily dismissed. The improvements in the physics of the atmospheric models used to determine the abundances, the success achieved in line-profile matching, and the self-consistency of the abundance determinations provide great credibility to the new lower abundances. However, solar models developed with the new abundances remain discrepant with helioseismic constraints, even with a variety of (often unjustified) changes to the input physics. Adjustments to the evolution of solar models, such as the early mass loss and low-Z accretion discussed here, show some promise but do not fully restore agreement. Any single adjustment to solar models does not fully resolve the problem. Combinations of changes might provide better agreement but seem contrived. A resolution to the solar abundance problem (or the solar model problem) remains elusive.

In the future, a more comprehensive exploration of parameter space, including opacity variations, different mass-loss or accretion prescriptions, diffusion, and perhaps even combinations of these effects could be useful. 
In particular, models with AGS05 abundances and a smaller amount of mass loss than explored here may provide a way to retain the core structure without completely destroying Li or creating too much $^3$He build-up. Further examination of the CO$^5$BOLD models with revised 
abundances for every element and opacity tables (including low-T 
opacities) adjusted for a new mixture would also be enlightening. 
\\ \\

\acknowledgements

The authors thank David Arnett, Alfio Bonanno, Piercarlo Bonifacio, Sarbani Basu, Joergen Christensen-Dalsgaard, Wick Haxton, Ross Rosenwald, Sylvaine Turck-Chi\`eze, and our anonymous referee for helpful discussions. We thank Scott Watson for code improvement and earlier versions of the AGS models.

\begin{comment}
***************

Notes:

To do list:\\
- add uncertainties to sound speed plot sound.eps\\
%\textbf{**for accretion: mention Castro; Zaatri; and Haxton (did it wrong - %no CZ mixing)**
\\
References to update:\\
- Lazrek et al 2006\\
- Nordlund 2009\\
- Arnett et al 2009\\

***************
\end{comment}

\bibliographystyle{aa}

\end{document}